\documentstyle[12pt,epsfig,psfrag]{article}

\newcommand{\kslash}{k \hspace{-0.21cm} /}
\newcommand{\qslash}{q \hspace{-0.2cm} /}
\newcommand{\pslash}{p \hspace{-0.18cm} /}
\newcommand{\pqslash}{p_{1} \hspace{-0.34cm} /}
\newcommand{\psslash}{p_{2} \hspace{-0.34cm} /}
\newcommand{\psmod}{|\vec{p}_2|}

\begin{document}

\title{\bf Gluon-Exchange in Spin-Dependent Fragmentation}      

\author{A.~Metz\footnote{e-mail: metza@tp2.ruhr-uni-bochum.de} \\[0.3cm]
{\it Institut f\"ur Theoretische Physik II,} \\
{\it Ruhr-Universit\"at Bochum, D-44780 Bochum, Germany}}

\date{\today}
\maketitle

\begin{abstract}
\noindent
The fragmentation of an unpolarized quark into a transversely
polarized spin-$\frac{1}{2}$ particle is studied in the framework of 
a simple model.
Special attention is payed to the gluon exchange which is incorporated 
in the gauge link of the fragmentation function, and which we model
by an abelian gauge field.
The transverse single spin asymmetries in $e^+ e^-$ annihilation and 
semi-inclusive deep-inelastic scattering are calculated in the
one-loop approximation.
For $e^+ e^-$ annihilation one finds a cancellation between
contributions from two on-shell intermediate states, which have no 
counterpart in deep-inelastic scattering.
As a consequence of this cancellation, the model predicts the same
spin asymmetry for both processes implying that, in the one-loop
approximation, the corresponding fragmentation function is universal.
\end{abstract}

\begin{figure}[t!]
\includegraphics[width=16.5cm]{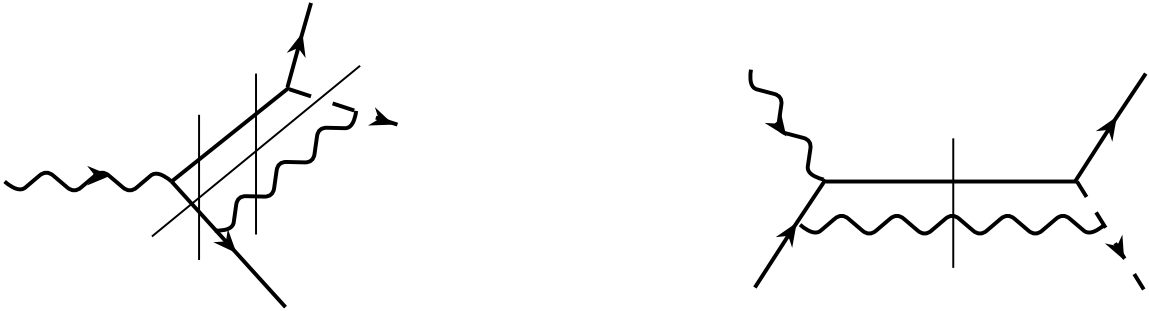} 
\caption{One-loop diagrams of fragmentation in $e^+ e^-$
 annihilation and semi-inclusive DIS. 
 The possible on-shell intermediate states are indicated by thin lines. 
\label{f:loop}}
\end{figure}

\noindent
Recently, considering inclusive deep-inelastic scattering (DIS), 
the influence of (Coulomb) gluon exchange between the struck quark 
and target spectators has been studied in detail~\cite{brodsky_02a}.
It has been emphasized that the rescattering of the struck quark
causes (additional) on-shell intermediate states in the forward
Compton amplitude, resulting in a shadowing contribution to the DIS 
cross section at leading twist.
In Feynman gauge, this shadowing effect is described by the gauge 
link (path-ordered exponential) appearing in the definiton of parton 
distributions~\cite{brodsky_02a}.
\\
Subsequently, the effect of rescattering has also been investigated in 
the case of semi-inclusive DIS~\cite{brodsky_02b}.
Using a simple model, it has been shown that a transverse 
single target-spin asymmetry arises from the interference between the
tree-level amplitude of the fragmentation process and the imaginary
part of the one-loop amplitude, where the latter describes the gluon 
exchange between the struck quark and the target system.
Afterwards, it has been demonstrated that the asymmetry
calculated in Ref.~\cite{brodsky_02b} is nothing else but a model for
the Sivers function including its gauge link~\cite{collins_02}.
The Sivers function, which belongs to the class of so-called
time-reversal odd (T-odd) and transverse momentum dependent 
($k_{\perp}$-dependent) parton densities, describes the distribution 
of unpolarized quarks in a transversely polarized 
target~\cite{sivers_90}.
Its existence requires a relative transverse momentum between the
target and the quark.
Initiated by the recent studies on the Sivers asymmetry,
the nontrivial question about the appropriate gauge link for 
$k_{\perp}$-dependent parton distributions in lightcone gauge has 
been addressed lately~\cite{ji_02,belitsky_02}.
\\
By considering the behaviour of the path-ordered exponential under
time-reversal a very interesting observation has been made in 
Ref.~\cite{collins_02}:
the Sivers asymmetry in semi-inlcusive DIS has the opposite sign
compared to the one in Drell-Yan, i.e., the Sivers function is
non-universal. 
This sign difference has been confirmed by an explicit 
model-calculation~\cite{brodsky_02c}.
Comparing DIS and Drell-Yan, in the meantime also for unpolarized 
scattering a violation of universality has been pointed out, provided 
that the cross section is kept differential in target-related 
particles~\cite{peigne_02}.
\\
In the present paper we study the universality of T-odd 
spin-dependent fragmentation functions.
At leading order in $1/Q$, where $Q$ denotes the hard scale of the 
process, two such objects exist for the fragmentation into a 
spin-$\frac{1}{2}$ hadron: the Collins function~\cite{collins_93} 
(fragmentation of a transversely polarized quark into an unpolarized 
hadron), and the Sivers-type fragmentation function~\cite{mulders_96} 
(fragmentation of an unpolarized quark into a transversely polarized 
spin-$\frac{1}{2}$ hadron).
It is important to notice that, in contrast to T-odd parton densities, 
these functions are non-vanishing in general, even if their gauge link 
is neglected~\cite{collins_93,bacchetta_01}.
Rescattering of hadrons in the fragmentation process can provide the
required imaginary part in the scattering amplitude.
\\
In the following, we explore the effect due to gluon exchange as
incorporated in the gauge link of the fragmentation functions.
We focus on the Sivers-type single spin asymmetry, which is 
of particular interest for $\Lambda$ production 
(see e.g.~\cite{anselmino_02}). 
We employ the simple model used in 
Refs.~\cite{brodsky_02b,brodsky_02c} to calculate the transverse 
spin asymmetry for $e^+ e^-$ annihilation and  semi-inclusive DIS 
at the one-loop level.
While for DIS only one on-shell intermediate state contributes to 
the asymmetry, there are three in $e^+ e^-$ annihilation. 
However, the contributions of two of them cancel out each other, and
the spin asymmetries for both processes are equal.
Therefore, the Sivers fragmentation function is universal at one-loop.
The same conclusion applies to the Collins function as well.
\begin{figure}[t!]
\includegraphics[width=16.5cm]{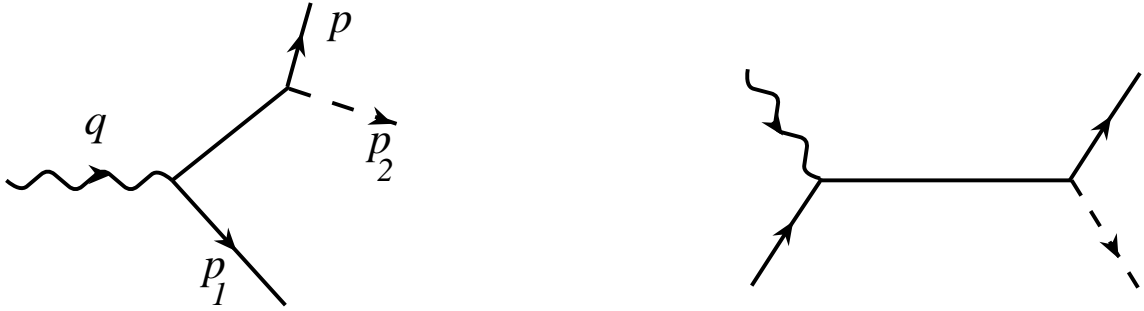} 
\caption{Tree-level diagrams of fragmentation in $e^+ e^-$
 annihilation and semi-inclusive DIS. 
 In both cases a quark fragments into a spin-$\frac{1}{2}$ hadron
 and a scalar remnant (dashed line). 
\label{f:tree}}
\end{figure}
\\[0.6cm]
\noindent
In Fig.~\ref{f:tree}, the tree-level diagrams of the two
fragmentation processes are displayed. 
For $e^+ e^-$ annihilation we consider the decay of a timelike virtual
photon into a $q\bar{q}$ pair, where the quark subsequently fragments
into a spin-$\frac{1}{2}$ hadron (in the following we frequently talk
about a proton) and a scalar remnant, i.e.,
\begin{equation}
 \gamma^{\ast}(q) \to \bar{q}(p_1,\lambda') + p(p,\lambda) + s(p_2) 
 \,. 
\end{equation}
The proton, including its polarization, is detected. 
The antiquark in the final state forms a jet.
Just as well one might consider the fragmentation of the antiquark
into an unpolarized hadron which, however, unnecessarily complicates 
the calculation even further.
To make the transition from $e^+ e^-$ annihilation to semi-inclusive
DIS one replaces the timelike photon by a spacelike one, and the 
outgoing antiquark by a quark in the initial state.
\\
The one-loop corrections are shown in Fig.~\ref{f:loop}. 
For $e^+ e^-$ annihilation (semi-inclusive DIS) a single photon is
exchanged between the remnant and the antiquark (initial quark).
These diagrams provide a simple model for the lowest order
contribution of the path-ordered exponential of the fragmentation
function.
(Actually, also the graph representing the one-loop correction 
to the vertex of the incoming photon in both processes is related 
to the path-ordered exponential.\footnote{The author thanks John
Collins for pointing this out.}
However, taking this diagram into account does not change any
conclusion of the present work.) 
Obviously, two cuts (on-shell quark and antiquark, as well as on-shell
antiquark and remnant) for $e^+ e^-$ annihilation have no
counterpart in semi-inclusive DIS.
Below we will demonstrate that these two on-shell states contribute to
the transverse spin asymmetry.
However, their contributions exactly cancel out each other.
The quark-photon cut in $e^+ e^-$ annihilation corresponds to the cut 
in DIS.
For the two processes the spin asymmetry due to the on-shell 
$q\gamma$ state is equal.
Particularly in the case of the $q\gamma$ cut we will only quote the 
final result for the asymmetry and present details of the calculation, 
which is interesting in its own, elsewhere~\cite{metz_02b}.
\\[0.6cm]
Before dealing with the model calculation, the kinematics is briefly
discussed.
We consider $e^+ e^-$ annihilation in the rest frame of the timelike 
photon.
The proton in the final state has no transverse momentum, and its 
minus-momentum is given by $zq^-$, where $q^-$ is the minus-momentum 
of the virtual photon.
We fix the plus-momentum of the antiquark according to 
$p_1^+ \approx q^+$.
The antiquark also carries a soft transverse momentum 
$-\vec{\Delta}_\perp$, implying that the fragmenting quark and the 
outgoing proton have a relative transverse momentum, which is 
necessary for the Sivers asymmetry. 
These requirements specify the kinematics:
\begin{eqnarray} \label{e:kin}
q & = & \bigg( Q \,, Q \,, \vec{0}_\perp \bigg) \,, 
\nonumber \\
p_1 & = & \bigg( Q \,, \frac{\vec{\Delta}_\perp^2 + m_q^2}{Q} \,, 
 -\vec{\Delta}_\perp \bigg) \,,
\nonumber \\
p & = & \bigg(\frac{M^2}{z Q} \,, z Q \,, \vec{0}_\perp \bigg) \,,
\nonumber \\
p_2 & = & \bigg( \frac{\vec{\Delta}_\perp^2 + m_s^2}{(1 - z) Q} 
 \,, (1 - z) Q \,,  \vec{\Delta}_\perp \bigg) \,.
\end{eqnarray}
For simplicity, we listed in~(\ref{e:kin}) always just the leading
terms.
Sometimes the $1/Q^2$ corrections of $p_1^+$ and $p_2^-$ are needed,
which can be obtained readily from 4-momentum conservation.
\\[0.6cm]
To calculate the Sivers-type fragmentation in $e^+ e^-$ annihilation 
the transverse component of the hadronic current has to be 
considered~\cite{boer_97}.
We define the various components of the current, depending on the 
helicities of the proton and the antiquark, via the invariant decay
amplitude $T$ according to
\begin{equation}
T(\lambda,\lambda') = \varepsilon_{\mu} J^{\mu}(\lambda,\lambda') \,,
\end{equation}
where $\varepsilon$ is the polarization vector of the virtual photon.
In the following we focus on the $x$-component $J^1$.
\\
In the model we are using (see Ref.~\cite{brodsky_02b}) the proton 
carries no electromagnetic charge.
Therefore, the charge of the fragmenting quark (denoted by $e_1$) and
the one of the remnant are equal.
The interaction between the quark, the proton, and the remnant is
described by a scalar vertex with the coupling constant $g$.
This yields for the diagram on the {\it lhs} in Fig.~\ref{f:tree} 
the current
\begin{eqnarray} \label{e:amptree}
J_{(0)}^1(\lambda,\lambda') & = & e_1 \, g \, 
 \frac{1}{s - m_q^2} \, 
 \bar{u}(p,\lambda) \, (\qslash - \pqslash + m_q) \, \gamma^1 \,
 v(p_1,\lambda') 
 \nonumber \\
 & = & e_1 \, g \, \frac{1 - z}{\sqrt{z}} \,
 \frac{Q}{\vec{\Delta}_{\perp}^2 + \tilde{m}^2} \,
 \bigg[ (\Delta^1 - i \lambda \Delta^2) \delta_{\lambda,-\lambda'}
 - \lambda \, \bigg( \frac{M}{z} + m_q \bigg) 
 \delta_{\lambda,\lambda'} \bigg] \,,
\\
& & {\rm with} \quad 
 \tilde{m}^2 = \frac{1}{z} \bigg( M^2 \, \frac{1 - z}{z} 
                  + m_s^2 - m_q^2 \, (1 - z) \bigg) \,,
\nonumber
\end{eqnarray}
where the lightfront helicity spinors of Ref.~\cite{lepage_80} 
have been employed to evaluate the matrix element.
We have also made use of the relation
\begin{equation} \label{e:sdef}
s - m_q^2 = (q - p_1)^2 - m_q^2 
= \frac{z}{1 - z} \, \Big( \vec{\Delta}_{\perp}^2 +\tilde{m}^2 \Big) \,,
\end{equation}
which connects the total energy $\sqrt{s}$ in the {\it cm}-frame of the 
outgoing proton and remnant with the variables $z$ and 
$\vec{\Delta}_{\perp}^2$.
\\[0.6cm]
In the next step the one-loop correction on the {\it lhs} 
in Fig.~\ref{f:loop} is included.
To calculate the single spin asymmetry only the imaginary part of this
diagram is important, where we focus here on the imaginary part caused 
by the on-shell $q\bar{q}$ intermediate state.
Generally, the imaginary part of a Feynman diagram is conveniently 
calculated by means of Cutkosky rules which determine the
discontinuity of a diagram.
Applying these rules the imaginary part of the one-loop graph is given
by 
\begin{eqnarray} \label{e:oneloopamp}
\lefteqn{{\rm Im}_{q \bar{q}} \, J_{(1)}^1 (\lambda,\lambda') 
 = \frac{1}{2 i} \, {\rm Disc}_{q \bar{q}} \, 
   J_{(1)}^1 (\lambda,\lambda')}
\nonumber \\
& = & - \frac{1}{2i} \, i \, (e_1)^3 \, g \, 
 \int \frac{d^4 k}{(2 \pi)^4} \, (-2 \pi i)^2 \, 
 \delta((p - k)^2 - m_q^2) \, 
 \delta((p - q - k)^2 - m_q^2)
\nonumber \\
& & \mbox{} \times
 \frac{\bar{u}(p,\lambda) \, (\pslash - \kslash + m_q) \, 
 \gamma^1 \, (\pslash - \qslash - \kslash + m_q) \,
 (\psslash - \kslash) \, v(p_1,\lambda')}
 {[k^2 - m_s^2 + i \epsilon] \, [(p_2 + k)^2 - \mu^2 + i \epsilon]}
 \,.
\end{eqnarray}
Note that we have assigned a mass $\mu$ to the gauge boson in order to avoid 
infrared singularities at intermediate steps of the calculation.
The same recipe has been used in the calculation of the target-spin
asymmetry~\cite{brodsky_02b,brodsky_02c}.
In the final result of the spin asymmetry the limit $\mu \to 0$ can 
be performed without encountering a divergence.
\\
The $\delta$-functions in~(\ref{e:oneloopamp}) are exploited to
perform the integrations over $k^+$ and $k^-$.
We rewrite them according to
\begin{eqnarray}
\delta((p - k)^2 - m_q^2) & = &
 \frac{1}{|k^- - p^-|} \, \delta \bigg( k^+ - p^+ -
 \frac{\vec{k}_{\perp}^2 + m_q^2}{k^- - p^-} \bigg) \,,
\\
\delta((p - q - k)^2 - m_q^2) & = &
 \frac{1}{Q} \, \delta(k^- - (p^- + p^+ - k^+ - Q)) \,,
\end{eqnarray}
and, hence, obtain
\begin{eqnarray} \label{e:delta}
\lefteqn{\int dk^+ dk^- \, \delta((p - k)^2 - m_q^2) \,
 \delta((p - q - k)^2 - m_q^2) \ldots}
 \nonumber \\
& = & \frac{1}{Q} \int dk^- \, \frac{1}{|k^- - p^-|} \,
\delta \bigg( k^- - p^- 
 + \frac{\vec{k}_{\perp}^2 + m_q^2}{k^- - p^-} +Q \bigg) \ldots 
 \bigg|_{\begin{array}{ccl}
         k^+ & = & p^+ 
                 + \frac{\vec{k}_{\perp}^2 + m_q^2}{k^- - p^-} 
         \end{array}}
 \nonumber \\
& = & \frac{1}{Q^2} \ldots 
 \bigg|_{\begin{array}{ccl}
         k^- & = & - (1 - z) Q \\
         k^+ & = & \frac{M^2}{z Q} 
                   - \frac{\vec{k}_{\perp}^2 + m_q^2}{Q}
         \end{array}} \,.
\end{eqnarray}
From the second line in Eq.~(\ref{e:delta}) one finds a quadratic 
equation for $k^-$.
Though this equation has two real solutions, only one of them
provides a leading twist contribution in the end.
To obtain the result in~(\ref{e:delta}) it also enters that, for the
leading twist part of the current, the transverse loop-momentum 
satisfies the condition $k_{\perp} \ll Q$.
\\
Before performing the $k_{\perp}$-integration, the matrix element in 
the numerator of Eq.~(\ref{e:oneloopamp}) is evaluated, keeping only
those terms that can contribute at leading order in the hard scale,
\begin{equation}
{\rm Im}_{q \bar{q}} \, J_{(1)}^1 (\lambda,\lambda') =
 \frac{(e_1)^3 \, g}{8 \pi^2} \, \frac{1 - z}{\sqrt{z}} \, Q
 \int d^2 \vec{k}_{\perp}
 \frac{(k^1 - i \lambda k^2) \, \delta_{\lambda,-\lambda'} 
 + \lambda \Big(\frac{M}{z} + m_q \Big) \, \delta_{\lambda,\lambda'}}
 {[\vec{k}_{\perp}^2 + \tilde{m}^2] \, 
  [(\vec{k}_{\perp} + \vec{\Delta}_{\perp})^2 + \mu^2]} \,.
\end{equation}
In order to carry out the $k_{\perp}$-integration it is convenient 
to combine the two factors in the denominator by means of the Feynman 
parameterization.
For instance, this allows one to write in the case of the scalar
integral:
\begin{eqnarray} \label{e:fpara}
\lefteqn{ \int d^2 \vec{k}_{\perp}
 \frac{1}{[\vec{k}_{\perp}^2 + \tilde{m}^2] \, 
  [(\vec{k}_{\perp} + \vec{\Delta}_{\perp})^2 + \mu^2]}}
\nonumber \\
& & = \int_0^1 d\alpha \int d^2 \vec{k}_{\perp}
\frac{1} {[\vec{k}_{\perp}^2 + \vec{\Delta}_\perp^2 \alpha (1 - \alpha) 
 + \mu^2 \alpha + \tilde{m}^2 (1 - \alpha)]^2} \,. 
\end{eqnarray}
The $k_{\perp}$-integration on the {\it rhs} in Eq.~(\ref{e:fpara})
can be performed easily. 
The situation for the vector integral (containing a $k_{\perp}$ 
in the numerator) is analogous.
\\ 
Combining the tree level expression for the hadronic current 
in~(\ref{e:amptree}) with the imaginary part at one loop, 
one finds the result
\begin{eqnarray}
\lefteqn{J^1 (\lambda,\lambda') =
 e_1 \, g \, \frac{1 - z}{\sqrt{z}} \, Q}
 \nonumber \\
& & \mbox{} \times \bigg[ (\Delta^1 - i \lambda \Delta^2) 
 \bigg( h - i \frac{(e_1)^2}{8\pi} \, g_2 \bigg) \,
 \delta_{\lambda,-\lambda'}
 - \lambda \bigg( \frac{M}{z} + m_q \bigg)
 \bigg( h - i \frac{(e_1)^2}{8\pi} \, g_1 \bigg) \,
 \delta_{\lambda,\lambda'} \bigg] \,,
 \\
& & {\rm with} \quad 
 h = \frac{1}{\vec{\Delta}_{\perp}^2 + \tilde{m}^2} \,,
 \nonumber \\
& & \hspace{1.1cm}
 g_1 = \int_0^1 d\alpha \, \frac{1}
 {\vec{\Delta}_\perp^2 \alpha (1 - \alpha) + \mu^2 \alpha +
  \tilde{m}^2 (1 - \alpha)} \,, 
 \nonumber \\
& & \hspace{1.1cm}
 g_2 = \int_0^1 d\alpha \, \frac{\alpha}
 {\vec{\Delta}_\perp^2 \alpha (1 - \alpha) + \mu^2 \alpha +
  \tilde{m}^2 (1 - \alpha)} \,.
 \nonumber
\end{eqnarray}
Here the reason for introducing a finite mass of the gauge boson
becomes very transparent.
The functions $g_1$ and $g_2$ are divergent in the limit 
$\mu \to 0$, since in this case the integrands diverge for 
$\alpha \to 1$.
\\[0.6cm]
The last step is the calculation of the transverse spin asymmetry
$\sigma_{pol} / \sigma_{unp}$, where the unpolarized and polarized 
(polarization along $x$-axes) cross sections are given according to
\begin{eqnarray}
\sigma_{unp} & \propto & \frac{1}{2} \sum_{\lambda,\lambda'} 
 J^1(\lambda,\lambda') \, \Big(J^1(\lambda,\lambda') \Big)^{\ast} \,,
\nonumber \\
\sigma_{pol} & \propto & \frac{1}{2} \sum_{\lambda'} 
 \Big[ J^1(s_x = \uparrow,\lambda') \, 
 \Big(J^1(s_x = \uparrow,\lambda') \Big)^{\ast} 
  - J^1(s_x = \downarrow,\lambda') \, 
 \Big(J^1(s_x = \downarrow,\lambda') \Big)^{\ast} \Big] \,.
\end{eqnarray}
Eventually, one obtains from the on-shell $q \bar{q}$ intermediate
state the following contribution to the transverse single spin 
asymmetry:
\begin{eqnarray} \label{e:qq}
{\cal A}_{x,q\bar{q}} & = & \frac{(e_1)^2}{8\pi} \, 
 \frac{2 \, \Big( \frac{M}{z} + m_q \Big) \, \Delta^2}
 {\Big( \frac{M}{z} + m_q \Big)^2 + \vec{\Delta}_{\perp}^2} \, 
 \frac{g_1 - g_2}{h}
\nonumber \\
& = & \frac{(e_1)^2}{8\pi} \, 
 \frac{2 \, \Big( \frac{M}{z} + m_q \Big) \, \Delta^2}
 {\Big( \frac{M}{z} + m_q \Big)^2 + \vec{\Delta}_{\perp}^2} \, 
 \frac{\vec{\Delta}_{\perp}^2 + \tilde{m}^2}
      {\vec{\Delta}_{\perp}^2} \, 
 \ln \, \frac{\vec{\Delta}_{\perp}^2 + \tilde{m}^2}
      {\tilde{m}^2} \,. 
\end{eqnarray}
As expected, the asymmetry is proportional to the $y$-component
(component perpendicular to the proton spin) of the transverse
momentum.
Note that the transition $\mu \to 0$ has been performed.
The difference $g_1 - g_2$ is finite in this limit.
\\[0.6cm]
For the on-shell antiquark and scalar remnant in the intermediate 
state the calculation is very similar to the previous case.
In particular, it turns out that the asymmetry caused by this cut 
cancels out the one due to the $q\bar{q}$ cut, i.e.
\begin{equation} \label{e:cancel}
{\cal A}_{x,\bar{q}s} = - \, {\cal A}_{x,q\bar{q}} \,. 
\end{equation}
The cancellation between the contributions of these two on-shell
states is not a peculiarity of our specific model, but should rather
hold in general~\cite{metz_02b}.
Given that the two discontinuities have different cut-thresholds 
the result (\ref{e:cancel}) appears a bit surprising.
However, this difference becomes unimportant in our calculation 
since in both cases one is above the threshold due to the
(asymptotically) large $Q^2$ considered here. 
In fact, a closer inspection shows that $Q^2 (1-z)$ needs to be large
compared to typical soft scales of the process.
\\[0.6cm]
While our calculation of the two cuts discussed above is technically 
similar to the one of the target-spin asymmetry performed in 
Refs.~\cite{brodsky_02b,brodsky_02c}, we had to treat the $q \gamma$ 
cut along different lines.
Here, merely the final result for the asymmetry due to the on-shell 
$q\gamma$ intermediate state is listed, and details will be presented 
elsewhere~\cite{metz_02b}.
One finds
\begin{eqnarray} \label{e:qgamma}
{\cal A}_{x,q\gamma} & = & - \frac{(e_1)^2}{8\pi} \, 
 \frac{2 \, \frac{M}{z} \, \Delta^2}
 {\Big( \frac{M}{z} \Big)^2 + \vec{\Delta}_{\perp}^2} \, 
 \frac{\vec{\Delta}_{\perp}^2 + \tilde{m}^2}
      {\vec{\Delta}_{\perp}^2} \, 
 \bigg[ \ln \frac{p_{20} - \psmod \cos \alpha}{m_s}
        + \cos \alpha \ln \frac{p_{20} + \psmod}{m_s} 
 \nonumber \\
 & &  \hspace{5.5cm}
        + \frac{1 - \cos^2 \alpha}{4 (1 - z)}
         \bigg( 1 - \frac{p_{20}}{\psmod} 
                \ln \frac{p_{20} + \psmod}{m_s} \bigg) \bigg] \,,
\end{eqnarray}
where we restricted ourselves to the case $m_q = 0$ in order to 
simplify the expression.
In Eq.~(\ref{e:qgamma}) the energy and the three-momentum of the 
remnant, as well as the scattering angle (angle between the
antiquark and the remnant) in the {\it cm}-frame of the proton 
and the remnant appear.
In terms of the variables $z$ and $\vec{\Delta}_{\perp}^2$ these 
quantities read
\begin{eqnarray}
p_{20} & = & \frac{1}{2 \sqrt{s}} \,
 \Big( s + m_s^2 - M^2 \Big) \,,
 \nonumber \\
\psmod & = & \frac{1}{2 \sqrt{s}} \,
 \sqrt{\Big( s - (m_s + M)^2 \Big) 
       \Big( s - (m_s - M)^2 \Big)} \,,
 \nonumber \\
\cos \alpha & = & \frac{1}{2 \sqrt{s} \, \psmod} \,
 \Big( (2z - 1) \, s + m_s^2 - M^2 \Big) \,,
\end{eqnarray}
with $\sqrt{s}$ as given in Eq.~(\ref{e:sdef}). 
By explicit calculation we have shown that for semi-inclusive DIS the
spin-asymmetry caused by the on-shell $q\gamma$ state coincides with the 
result in~(\ref{e:qgamma}).\footnote{The amplitudes for both
processes are connected by a crossing relation.
In order to obtain the correct crossing behaviour it is mandatory that 
the vertices of the gauge boson with the quark and the antiquark have the 
same sign.
Using this rule we can confirm the reversed sign of the target-spin 
asymmetry for  Drell-Yan in comparison to the one for 
DIS~\cite{collins_02,brodsky_02c}.} 
Therefore, the total transverse spin asymmetries in $e^+ e^-$
annihilation and in semi-inclusive DIS are equal, and the Sivers
fragmentation function is universal in the one-loop model.
The same conclusion holds for the Collins function as well since the 
$q\gamma$ cut leads, for both processes, to the same imaginary part for
the four independent helicity amplitudes.
\\[0.6cm]
In summary, we have investigated the time-reversal odd fragmentation 
of an unpolarized quark into a transversely polarized
spin-$\frac{1}{2}$ hadron.
A one-loop approach with photon exchange has been used as a simple model
for the gluon exchange incorporated in the gauge link of the
fragmentation function.
One finds essentially two results by comparing the transverse single 
spin asymmetries in $e^+ e^-$ annihilation and semi-inclusive DIS:
firstly, in $e^+ e^-$ annihilation on-shell intermediate states exist 
which have no counterpart in DIS.
However, the contributions of these on-shell states to the spin 
asymmetry cancel out each other.
Secondly, the asymmetry caused by the one on-shell state, which
$e^+ e^-$ annihilation and semi-inclusive DIS have in common, is equal 
for both processes.
Therefore, the Sivers fragmentation function is universal.
The same conclusion holds as well for the Collins function which plays 
an important role in measurements of the transversity distribution of 
the nucleon.
It remains to be seen if the universality of time-reversal odd
fragmentation functions survives once higher order gluon-exchange
is taken into account.
\\
Finally, we repeat that a one-loop calculation does not provide a 
sign change for the Sivers fragmentation function, although the gauge 
link of an incoming fermion is replaced by the one of the
corresponding antifermion in the final state when making the
transition from semi-inclusive DIS to $e^+ e^-$ annihilation.
This is in contrast to time-reversal odd parton 
distributions~\cite{collins_02,brodsky_02c}.
\\[0.6cm]
\noindent
{\bf Acknowledgements:} I am very grateful to John Collins for drawing
my attention to the fact that in $e^+ e^-$ annihilation, compared to 
semi-inclusive DIS, two additional on-shell intermediate states may 
contribute to the transverse spin asymmetry. 
I also thank St\'ephane Peign\'e for a correspondence. 
This work has been supported by the Alexander von Humboldt Foundation.



\begin{thebibliography}{99}

\bibitem{brodsky_02a}
S.J.~Brodsky, P.~Hoyer, N.~Marchal, S.~Peign\'e, and F.~Sannino,
 Phys. Rev. D {\bf 65}, 114025 (2002).
\bibitem{brodsky_02b}
S.J.~Brodsky, D.S.~Hwang, and I.~Schmidt, 
 Phys. Lett. B {\bf 530}, 99 (2002).
\bibitem{collins_02}
J.C.~Collins,
 Phys. Lett. B {\bf 536}, 43 (2002).
\bibitem{sivers_90}
D.W.~Sivers,
 Phys. Rev. D {\bf 41}, 83 (1990); 
 Phys. Rev. D {\bf 43}, 261 (1991).
\bibitem{ji_02}
X.~Ji and F.~Yuan,
 Phys. Lett. B {\bf 543}, 66 (2002).
\bibitem{belitsky_02}
A.V.~Belitsky, X.~Ji, and F.~Yuan,
 hep-ph/0208038.
\bibitem{brodsky_02c}
S.J.~Brodsky, D.S.~Hwang, and I.~Schmidt, 
 Nucl. Phys. {\bf B642}, 344 (2002). 
\bibitem{peigne_02}
S. Peign\'e,
 hep-ph/0206138.
\bibitem{collins_93}
J.C.~Collins,
 Nucl. Phys. {\bf B396}, 161 (1993). 
\bibitem{mulders_96}
P.J.~Mulders and R.D.~Tangerman,
 Nucl. Phys. {\bf B461}, 197 (1996); 
 Nucl. Phys. {\bf B484}, 538(E) (1997).
\bibitem{bacchetta_01}
A.~Bacchetta, R.~Kundu, A.~Metz, and P.J.~Mulders,
 Phys. Lett. B {\bf 506}, 155 (2001);
 Phys. Rev. D {\bf 65}, 094021 (2002).
\bibitem{anselmino_02}
M.~Anselmino, D.~Boer, U.~D'Alesio, and F.~Murgia,
 Phys. Rev. D {\bf 65}, 114014 (2002). 
\bibitem{metz_02b}
A.~Metz,
 in preparation.
\bibitem{boer_97}
D.~Boer, R.~Jakob, and P.J.~Mulders,
 Nucl. Phys. {\bf B504}, 345 (1997). 
\bibitem{lepage_80}
G.P.~Lepage and S.J.~Brodsky, 
 Phys. Rev. D {\bf 22}, 2157 (1980).
\end{thebibliography}
\end{document}